\documentclass[3p,times]{elsarticle}

\usepackage{xspace}
\usepackage[T1]{fontenc}
\usepackage[scaled]{beramono}
\usepackage{listings}
\usepackage{tikz}
\usepackage{dingbat}
\usepackage{amssymb}
\usepackage{amsmath}
\usepackage{bbding}
\usepackage{pbox}
\usepackage{scrextend}
\usepackage{hyperref}
\usepackage{multirow}

\def\reff@jnl#1{{\rm#1\/}}
\def\aj{\reff@jnl{AJ}}         
\def\araa{\reff@jnl{ARA\&A}}      
\def\apj{\reff@jnl{ApJ}}        
\def\apjl{\reff@jnl{ApJ}}        
\def\apjs{\reff@jnl{ApJS}}       
\def\aap{\reff@jnl{A\&A}}        
\def\aapr{\reff@jnl{A\&A~Rev.}}     
\def\aaps{\reff@jnl{A\&AS}}       
\def\mnras{\reff@jnl{MNRAS}}      
\def\physrep{\reff@jnl{Physics Reports}}
\def\prd{\reff@jnl{Phys.Rev.D}}     
\def\prl{\reff@jnl{Phys.Rev.Lett}}   
\def\pasp{\reff@jnl{PASP}}       
\def\pasj{\reff@jnl{PASJ}}       
\def\nat{\reff@jnl{Nature}}       
\def\jcap{\reff@jnl{JCAP}}   
\def\memsai{\reff@jnl{MemSAI}} 
\def\na{\reff@jnl{New Astronomy}}       
\def\procspie{\reff@jnl{SPIE}}       
\def\pasa{\reff@jnl{PASA}}       

\newcommand{\tensorflow}{\texttt{Tensorflow}\xspace}
\newcommand{\unet}{\texttt{U-Net}\xspace}
\newcommand{\hide}{\texttt{HIDE}\xspace}
\newcommand{\seek}{\texttt{SEEK}\xspace}

\begin{document}

\begin{frontmatter}

\title{Radio frequency interference mitigation using deep convolutional neural
networks}

\author[eth1]{Jo\"el Akeret\corref{cor1}}
\ead{joel.akeret@phys.ethz.ch}

\author[eth1]{Chihway Chang}
\ead{chihway.chang@phys.ethz.ch}

\author[eth2]{Aurelien Lucchi}
\ead{aurelien.lucchi@inf.ethz.ch}

\author[eth1]{Alexandre Refregier}

\cortext[cor1]{Corresponding author}

\address[eth1]{ETH Zurich, Institute for Astronomy, Department of Physics, Wolfgang Pauli Strasse 27, 8093 Zurich, Switzerland}
\address[eth2]{ETH Zurich, Data Analytics Lab, Department of Computer Science, Universitaetstrasse 6, 8092 Zurich, Switzerland}
\begin{abstract}

We propose a novel approach for mitigating radio frequency interference (RFI) signals in radio data 
using the latest advances in deep learning. We employ a special type of Convolutional Neural 
Network, the \unet, that enables the classification of clean signal and RFI
signatures in 2D time-ordered data acquired from a radio telescope. We train and
assess the performance of this network using the \hide \& \seek radio data
simulation and processing packages, as well as early Science
Verification data acquired with the 7m single-dish telescope at the Bleien
Observatory.
We find that our \unet implementation is showing competitive accuracy
to classical RFI mitigation algorithms such as \seek's \textsc{SumThreshold} implementation. We publish our \unet software package on GitHub under GPLv3
license.

\end{abstract}


\end{frontmatter}

\section{Introduction}
\label{sec:introduction}

The radio band is becoming one of the most promising wavelength windows for cosmology. In 
particular, observations of the 21 cm neutral hydrogen line allows us to probe the high-redshift 
Universe, which is not easily accessible with other wavelengths \citep{pritchard2012cm}. In 
addition, radio band data 
provides important information for foreground studies of cosmic microwave background, and 
also Galactic astronomy \citep{chang2016integrated}. Ongoing and future
experiments such as LOFAR \citep{vanHaarlem2013}, GMRT \citep{paciga2013simulation},
PAPER \citep{ali2015paper}, CHIME \citep{bandura2014canadian}, BINGO
\citep{battye2013h, battye2012bingo}, HERA \citep{pober2014next}, Tianlai
\citep{chen2012tianlai}, and the SKA \citep{mellema2015hi} aim to carry out
wide-field surveys in the radio band that cover large portions of the sky.

One of the main challenges in all these surveys is the radio frequency interference 
(RFI) contamination to the data \cite{offringa2010post}. RFI can originate
from a wide variety of human produced sources such as satellites (GPS,
geostationary, TV etc.), cell phones, and air traffic communication. Different sources of RFI display different frequency and time-dependencies, 
causing the overall RFI signal to be complex and difficult to
model \cite{fridman2001rfi}. If the RFI signal is strong and mixed
with the astronomical signal of interest, the data cannot be used and will need to be masked. 

To minimize the RFI contamination to data, radio telescopes are normally built in remote 
locations that are protected against major human-made emission sources. Some
level of hardware improvement such as ground-shielding and band-pass filters can also reduce the 
input of RFI. However, in almost all situations, RFI masking in the analysis
software will still be needed.

The goal of any RFI masking algorithm is to minimize the amount of data lost
while ensuring low  RFI contamination. This procedure typically relies on the common assumption that the morphological characteristics of RFI in the 2D plane of time and frequency (the raw data format of standard spectrometers) are different from that of astronomical signals. Astronomical signals are usually broad-band and 
vary smoothly over long time-scales, while RFI appears as high-intensity pixels localized in the 
time-frequency plane or is sometimes also periodic in time. Existing RFI
mitigation algorithms typically fall into three categories. The first category attempts to identify the characteristics of RFI through linear methods such as Singular Vector Decomposition (SVD)
\citep{offringa2010post} or Principle Component Analysis (PCA) 
\citep{zhao2013windsat}. These methods work well if the RFI pattern exhibits a
repeated pattern over time and frequency, but cannot handle with more stochastic
signals such as the ones caused by irregular satellites. The second category uses threshold-based algorithms such as \textsc{cumsum} \citep{baan2004radio} and \textsc{SumThreshold} \citep{offringa2010post}, where the RFI is defined as pixels above some threshold in the smoothed 2D time-frequency plane. Despite their simplicity, 
these methods are fairly reliable and can be quite effective. In particular, \textsc{SumThreshold} is 
the most widely used algorithm in existing radio data processing pipelines \citep{offringa2010post, 
offringa2010lofar, peck2013serpent, akeret2016hide}. The third category uses
traditional supervised machine-learning techniques such as K-nearest neighbour
and Gaussian mixture models to cluster RFI signals \citep{wolfaardt2016machine}.
For these methods to achieve a sufficient classification accuracy, a careful
feature selection process has to be performed prior to the application.
While these three classes of methods have encountered a significant success in
astronomy, somewhat more advanced techniques in machine learning have not been explored.

One approach that has shown promising results in the area of machine learning
are deep neural networks. In the recent years, they outperformed
state-of-the-art techniques in various classification tasks such as biomedical image segmentation \citep{ronneberger2015u} or natural language processing \citep{collobert2011natural}.
Although the concept of artificial neural network has
been around for many years, their current preeminence can be mostly attributed
to recent advances in customized hardware (especially GPUs) as well as the
development of open source deep learning software packages\footnote{We here
will be using \tensorflow, a recent deep learning framework released by Google.}.

A particular successful type of network is the convolutional neural network
(CNN)\cite[e.g.][]{krizhevsky2012imagenet, collobert2008unified}. Typically, CNNs have been used to
detect objects in images (without having any exact prior knowledge of
where the object appears in the image). These networks have also recently been
extended to the problem of image segmentation, for which a class label is
assigned to each pixel in an input image. One example of this segmentation
network is the \unet\cite{ronneberger2015u}. In this paper we apply this type of
CNN to identify and mitigate RFI in time-ordered-data (TOD) of a single-dish
radio telescope. To the best of our knowledge, this is the first application of deep learning techniques to this class of problem.

This paper is organized as follows. In Section \ref{sec:architecture} we describe the basic architecture 
and design of the \unet. In Section \ref{sec:results}, we apply the CNN to mitigate RFI on data taken at 
the Bleien Observatory. This includes a discussion of the performance of the CNN both on simulated and 
observed data. We then conclude in Sections \ref{sec:conclusion}. Information for downloading and installing 
our implementation of the \unet package is described in \ref{sec:distribution}. In \ref{sec:usage} we explain 
how to use the package.

\section{Proposed approach}
\label{sec:architecture}

\subsection{Network architecture}

The \unet\cite{ronneberger2015u} extends the architecture of conventional CNN's.
Typically, CNN's extract image features by repeatedly applying convolutions on
the input image followed by an activation function and a downsampling operation.
These nested operations let the network build a conceptual hierarchy of the
content present in the training images. Some similarities can be drawn to the
human visual system where the early layers extract small, localized features
such as edges while deeper layer combines these extracted edges into more
complex representations. Note that the downsampling operations present in a CNN
lead to a contraction of the information flowing through the network. This makes
conventional CNNs not well suited for image segmentation.

\begin{figure}
\begin{center}
\includegraphics[width=0.8\linewidth]{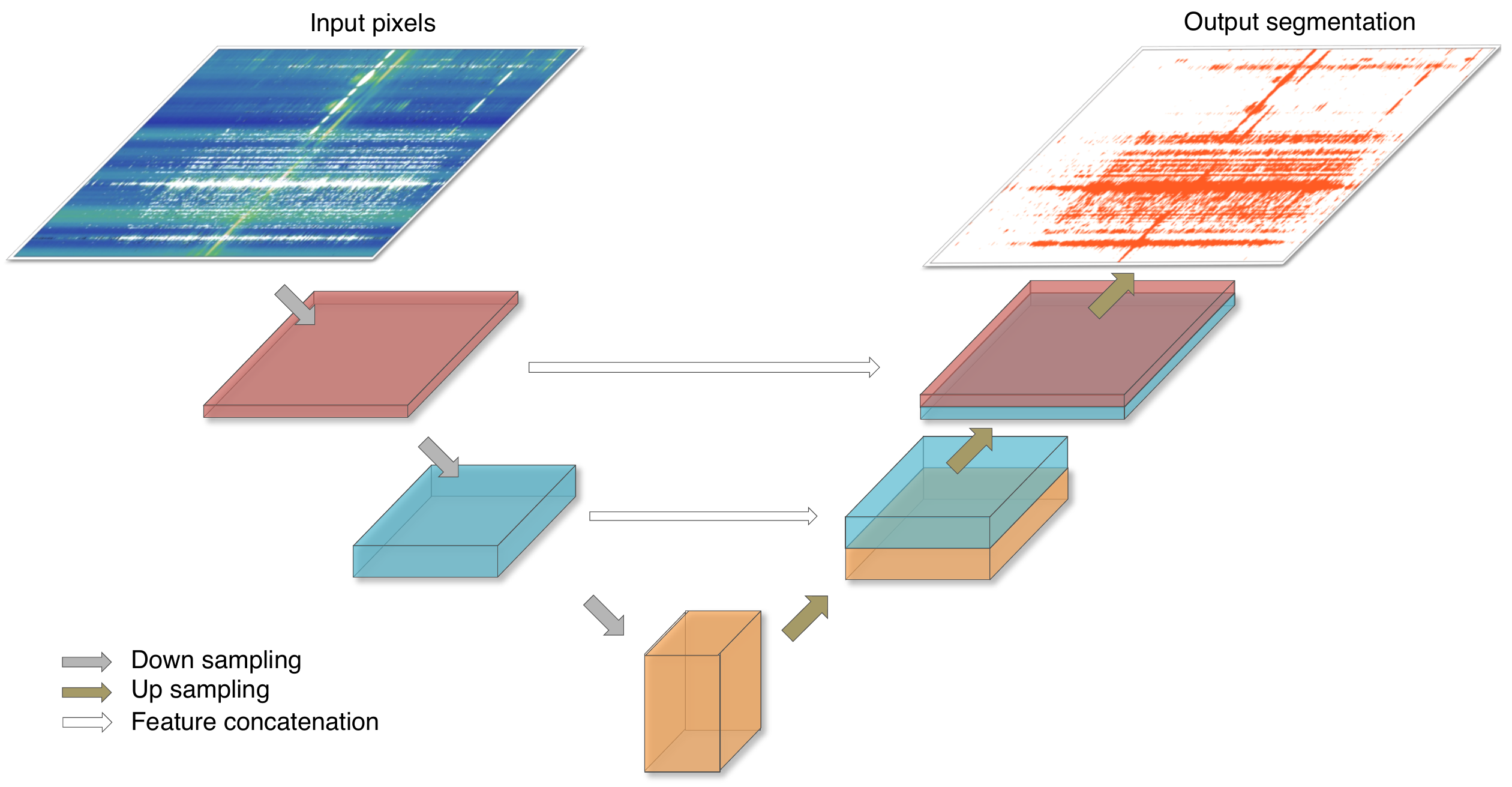}
\caption{Conceptional architecture of the \unet with a layer depth of 3. The pixel
information of the input image is contracted in the downsampling path. The extracted features are
then propagated to the higher layers in the upsampling path leading to the
output segmentation. The heights of the boxes represent the number of extracted
features and the white arrows show the feature concatenation.}
\label{fig:architecture}
\end{center}
\end{figure}

Instead of relying on a traditional architecture, the \unet extends the
contracting path of a CNN by a symmetric expansive path. As shown in Figure
\ref{fig:architecture}, the information on the extracted complex features
(orange box) from the pooling path are propagated to the higher layers by
several upsampling operations. The downsampling path followed by the upsampling path resembles a U-shape leading to the name of this network architecture.

We have reimplemented the original \unet\cite{ronneberger2015u}, written in
Caffe, with the open source library \tensorflow following its exact
architecture. Our \tensorflow \unet implementation is written in Python with maximal
flexibility in mind. The package is published on
GitHub\footnote{\url{http://github.com/jakeret/tf_unet}} under GPLv3 license
and can be used for various classification tasks (see \ref{sec:distribution}
and \ref{sec:usage} for installation instructions and usage examples).
In the contracting path we apply in each layer two consecutive unpadded convolutions both followed by a
rectifier linear unit (ReLU) activation and a $2 \times 2$ max pooling downsampling
operation. At each layer we double the number of extracted features. In the
expansive path we replace the max pooling by an up-convolution that halves the
number of features from the previous layer and concatenate the result with the
features from the corresponding contraction layer. Finally, we apply a $1 \times 1$
convolution to map the features from the last layer to the number of class
labels i.e. to a binary decision if a pixel is contaminated or not. To obtain
the probability of a pixel to belong to a certain class we convert the resulting
output map with a pixel-wise soft-max layer. The RFI mitigation is done by
inputting the TOD and applying a threshold on the predicted probability of each pixel to be contaminated with RFI.

\subsection{Training the network}

We train the parameters of the \unet using the early Science Verification data
acquired at the Bleien Observatory \cite{chang2016integrated}. 
This data set was collected using a 7m single-dish telescope operating in 
drift-scan mode with a frequency range of 990 -- 1260 MHz.
We have processed the data with the \hide \& \seek
radio data processing pipelines described in \cite{akeret2016hide}. The pipeline
employs the \textsc{SumThreshold} algorithm to mask pixels contaminated with
RFI. \textsc{SumThreshold} is a widely used iterative algorithm
that is gradually building a mask to flag the unwanted signal. It follows the
underlying assumption that the astronomical signal is relatively smooth, both,
in time and frequency direction. While RFI signal exhibit patterns with sharp
edges. The algorithm gradually improves a model of the astronomical signal and
masks values lying above a certain threshold after subtracting this model from
the data. It starts with localized, strong RFI bursts and extends the mask by
gradually analyzing the neighboring pixels \citep{akeret2016hide}. The parameters 
we adopt for the \textsc{SumThreshold} algorithm here are based on the procedure
developed in
\cite{akeret2016hide}. We use the \textsc{SumThreshold} 
mask as ground truth to train the neural network as well as to evaluate the
performance of the network on a separate test set. We note, however, that the
RFI mask produced by \textsc{SumThreshold} is not perfect. It has a high
false-positive-rate i.e. many pixels are incorrectly flagged as RFI. 
Some RFI detection pipelines have refined this technique, e.g. by using a scale
invariant dilation operation\cite{offringa2012morphological}. This can improve
the flagging performance of the algorithm. However, we demonstrate in this
paper that our \unet model is robust to this noise in the ground truth and is
capable of correctly distinguishing between non-contaminated and contaminated pixels.

We explore the effects of various parameters on the classification performance
and processing time. Here we report the effect of the parameters that most
influence the performance such as the depth of the network (i.e. the number of
layers), the number of features extracted in the first layer, and the size
of the convolution kernels. We optimize a cross-entropy loss function to train
the network parameters using a momentum-based stochastic gradient decent with an
exponentially decaying learning rate with an initial value of 0.2.
We initialize the weights of the network using a truncated normal distribution
following the recommendation for the standard deviation in
\cite{ronneberger2015u}. We train each network for 100 epochs each with a
training mini-batch size of 32 on one TOD image with a resolution of $276 \times
600$ pixels.
For all configurations the loss function had reached a minima and remained
stable. In order to avoid overfitting we use dropout layers
\cite{srivastava2014dropout} with a probability of 0.5 in all convolutions
combined with an L2 regularizer of strength equal to $10^{-3}$. We train
our networks using a NVIDIA Kepler K20 GPU.
 
\section{Experimental results}
\label{sec:results}

\begin{figure}[t]
\begin{center}
\includegraphics[width=0.49\linewidth]{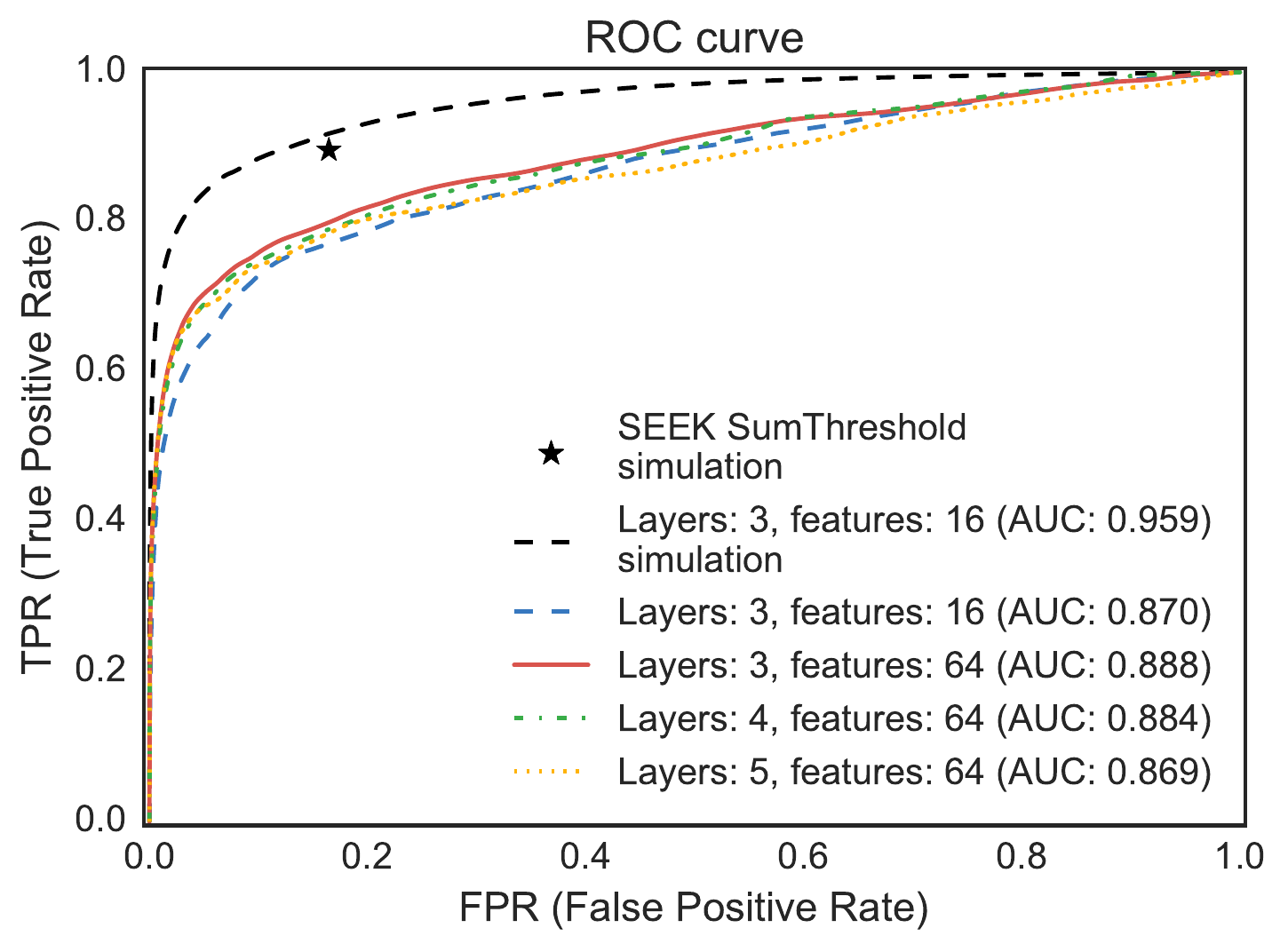}
\includegraphics[width=0.49\linewidth]{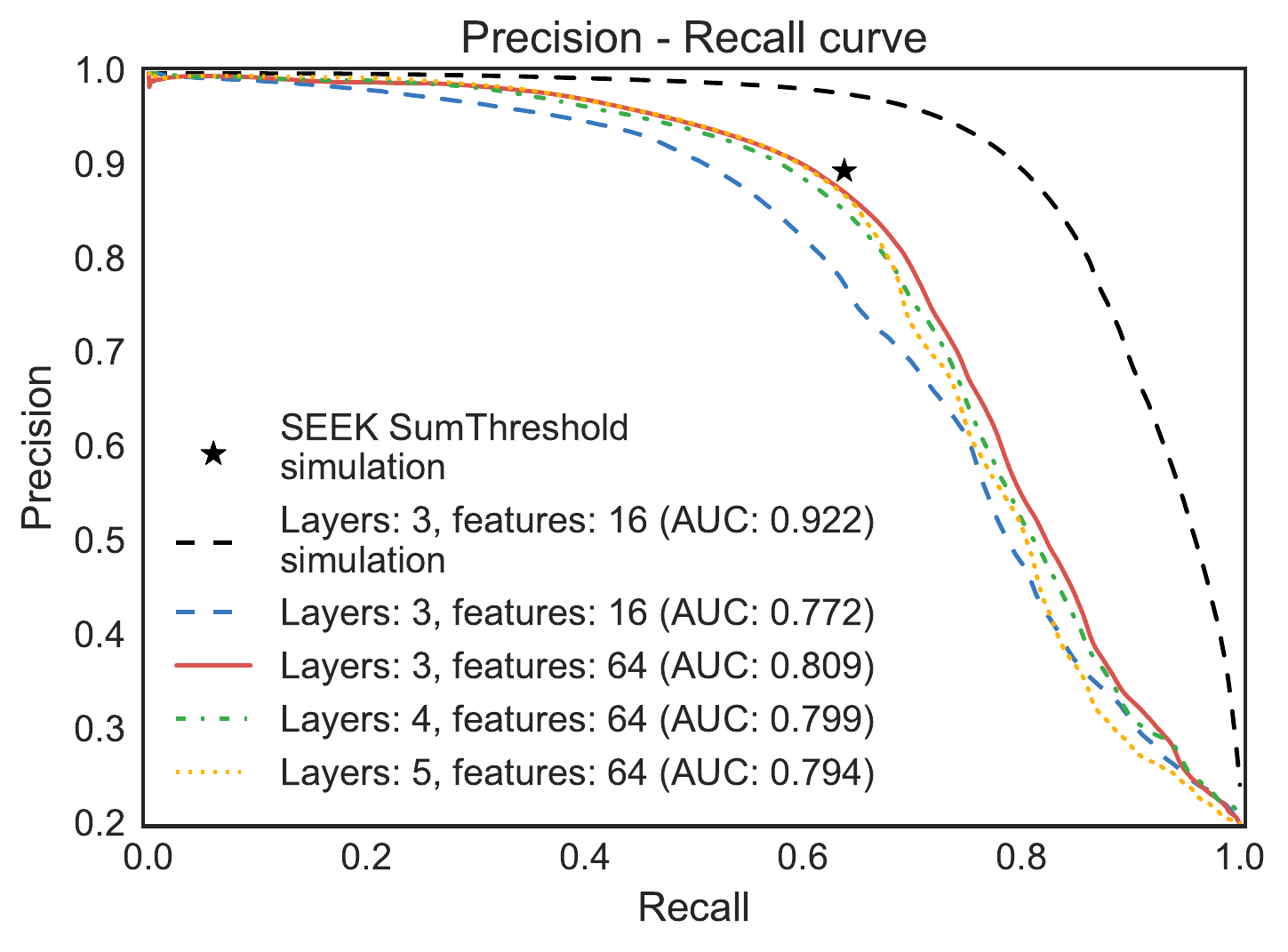}
\caption{The left panel shows the receiver operating characteristic (ROC) curve for different \unet configurations. 
The right panel depicts the precision-recall curve for the same architectures.
The black dashed line shows the performance of the \unet on simulated data sets
while the other lines are showing the results on real data. The
stars denote the performance of \seek's \textsc{SumThreshold} on
the simulated data set.}
\label{fig:performance}
\end{center}
\end{figure}

To assess if a CNN can be used for the RFI mitigation task described in the
previous section, we first apply our \unet implementation to simulated radio
data. We used the publicly available \hide package to simulate radio data
contaminated by RFI. The simulation used here reflect the drift-scan
data of the early Science Verification survey taken at the Bleien Observatory.
\hide simulates the noise effects of the instrument and the atmosphere on the data.
Furthermore, elevation-dependent signal variations, such as side-lobe
ground-pickup are also taken into account. The RFI model employed in the package
simulates a simple but well understood interference pattern. Each RFI burst is
defined by the same profile. The amplitude and the rate of the burst is randomly
sampled such that the expected data loss of the observation site is matched.
\cite{akeret2016hide, chang2016integrated}. We have access to the perfect ground
truth of this simulated data since we know for each pixel if it has been perturbed by
simulated RFI. This simplified the training procedure and allowed us to better
quantify the performance of the network. We find that our \unet implementation
performs very well in identifying RFI pixels. As shown in Figure
\ref{fig:performance} the \unet achieves an Area-Under-the-Curve (AUC) score of
$\sim$0.96 and $\sim$0.92 for the receiver operating characteristic (ROC) and
the Precision-Recall curve respectively (black dashed line). The black stars
denotes the performance of \seek's \textsc{SumThreshold} on the same simulated
data set. We note that \seek is achieving a comparable performance in the ROC
metric but has a lower precision and recall than our \unet. The \unet achieves
a maximum $F_1$ score\footnote{The $F_1$ score can be seen as the weighted mean of precision and recall.
$0<F_1<1$ and a larger $F_1$ score indicates better classification performance.} of $\sim$0.85 compared to $\sim$0.75 for
\seek's \textsc{SumThreshold} algorithm.

In a second stage, we use the CNN trained on simulations to mitigate RFI in data
taken at the Bleien Observatory. As described in \cite{akeret2016hide} the RFI
simulation in \hide is relatively simple. Therefore the trained \unet struggled
to detect long lasting broadband RFI signatures in real data, such as satellite
emissions since they were not simulated in \hide at all.

\begin{figure}
\begin{center}
\includegraphics[height=3.7cm,keepaspectratio]{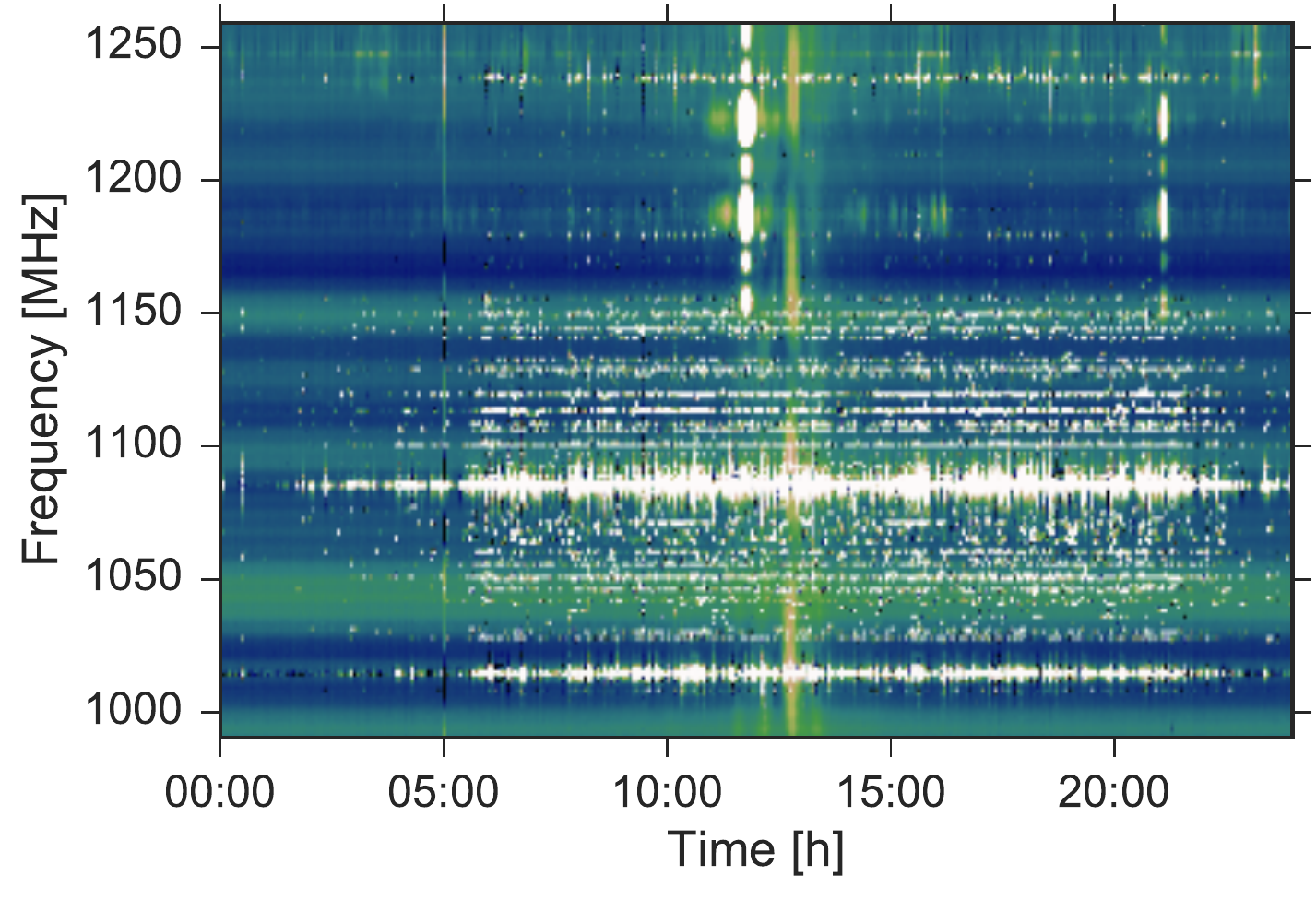}
\includegraphics[height=3.7cm,keepaspectratio]{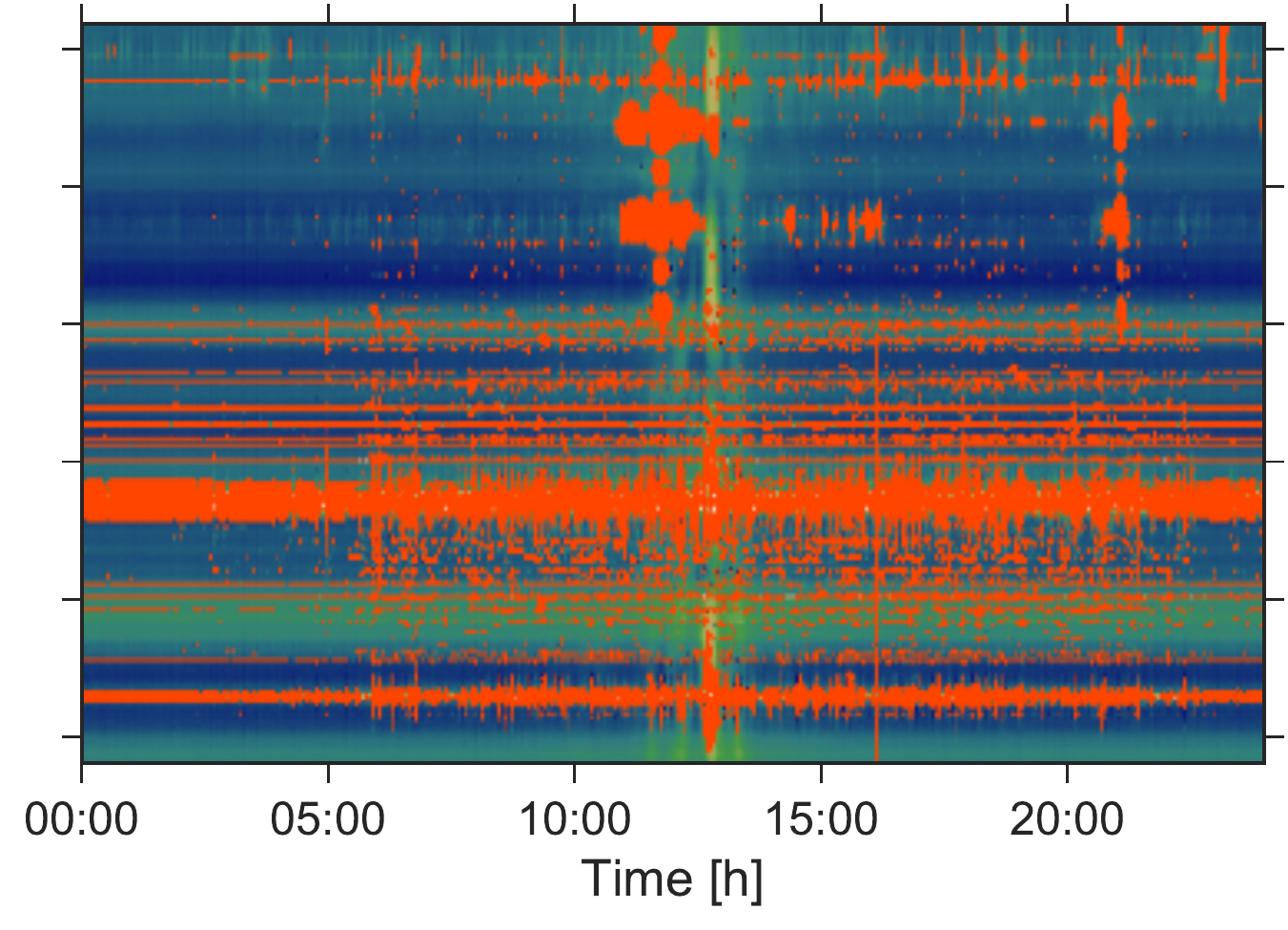}
\includegraphics[height=3.7cm,keepaspectratio]{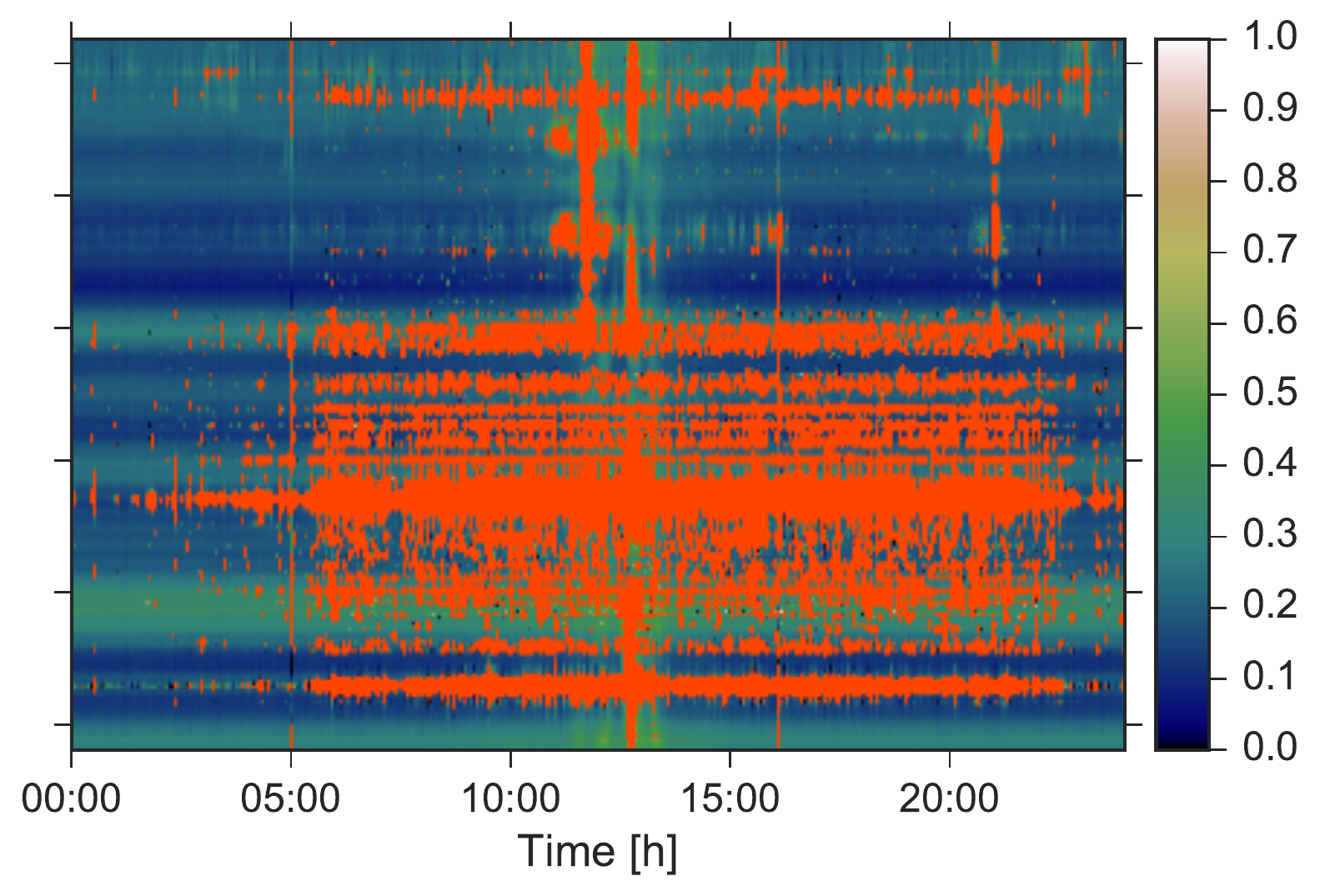}
\caption{The left panel displays 24 hours of observed TOD from the Bleien
Observatory.
The broadband RFI contamination mainly comes from the nearby airport and is visible in the 1025--1150 MHz frequency band. The TOD 
also demonstrates the variation in the RFI level between day and night as the amount of RFI clearly 
increased at around 6:00 am and decreased at 11:00 pm. The central panel shows the same TOD 
overlaid (orange) with the RFI mask obtained from \seek's \textsc{SumThreshold}. The right panel 
displays the RFI mask from our \unet with 3 layers and 64 features. }
\label{fig:tod}
\end{center}
\end{figure}

To achieve a better classification result on the observed data set we have processed the data with 
the \seek data processing pipeline to obtain a RFI mask for the TOD derived from the 
\textsc{SumThreshold} algorithm. Compared to the perfect ground truth mask from the simulation, this 
\textsc{SumThreshold} mask contains incorrectly masked pixels and RFI pixels that were not detected. 
The left panel of Figure \ref{fig:tod} shows an example of a set of rescaled, observed TOD 
used for the performance verification. We observe that the data is strongly contaminated by narrow and 
broadband RFI. For example, the extended burst between 1150 and 1250 MHz at
around 11\textsc{am} can be attributed to a satellite passing through the
telescope beam.
The central panel displays the same dataset overlaid with \seek's \textsc{SumThreshold} mask. We 
find that the pipeline captures most of the RFI signal, but part of the uncontaminated data is incorrectly 
masked, e.g. in the range 1070 to 1100 MHz between 0 and 5\textsc{am}. 
The incorrect masking is mainly associated with the parameter setting of the mask dilation 
and smoothing. Defining these parameters is particularly challenging because the characteristics 
of the RFI in day time and night time are significantly different and its is hard for a single set of 
parameters to satisfy both. Here, for example, if we turn down the mask dilation to recover the clean 
data at night, the mask during the day will be sub-optimal

The left panel of Figure \ref{fig:performance} shows the ROC curve for different \unet configurations. The 
x-axis displays the ratio of pixels incorrectly labeled as RFI while the y-axis represents the ratio of correctly 
masked pixels. The right panel shows the precision-recall curve for the same
network configurations. The black dashed lines show the performance
for the network on simulated data. The other curves display the result on
the observed data sets. The colored lines in these plots show that, beyond a
certain network complexity, further increase in the number of layers or features does not
significantly improve the performance of the network. We find that using a
\unet with 3 layers and 64 features (red solid line) provides a good balance
in terms of prediction performance and computational cost. The
network can be trained in a few hours on a modern GPU. The RFI detection
throughput with this trained network is approximately 11.6 GB/h/GPU. The AUC
score is lower on the observed than on the simulation data set as expected. This can be attributed to the increased complexity of the data and more importantly, the imperfect ground truth.

The right panel of Figure \ref{fig:tod} shows the observed TOD overlaid with the mask obtained 
from our \unet with 3 layers and 64 features. The CNN captures narrow and
broadband RFI signatures very well such that the mask resembles \seek's
\textsc{SumThreshold} output. The network has correctly learned the
signatures of RFI signals such that it can compensate for the incorrectly
masked pixel of the imperfect ground truth. 
Compared to the \textsc{SumThreshold} mask, we see that the power of 
the CNN approach is that it can automatically learn the local characteristics of the 
data in the time-frequency plane in a flexible and natural way.

\section{Conclusion}
\label{sec:conclusion}

Large surveys in the radio wavelength, especially of the 21cm neutral hydrogen line, is emerging as one 
of the most promising probes for cosmology. The resulting data will allow us to
map out the structure of the Universe over a wide period of cosmic time and reveal parts of the Universe that is not accessible via other wavelengths. 
In the coming decades, we expect a very large amount of radio data to be collected and analyzed. One of 
the major challenges in processing radio data is the masking of radio frequency interference (RFI). The 
complex nature of RFI in both the time and frequency domain means that designing a general automated algorithm for RFI 
masking is a challenging task. The most popular existing method relies
on thresholding algorithms, which require tuning of the threshold parameters. In
some RFI masking pipelines this can be done automatically but for optimal
performance further, manual tuning is typically necessary. The false-positive
rate achieved by these algorithms often also depend on the RFI environment at
the telescope.

In this paper, we presented a novel approach to RFI mitigation which makes use
of recent advances in deep learning. We adopt a special architecture of a
Convolutional Neural Network (CNN) that has proven to be very successful for
various classification problems. This \unet learns a set of features extracted
from the input time-ordered-data (TOD) from a radio telescope in order to
distinguish between the astronomical signal and the diverse and complex RFI
signatures.
Our \unet is implemented with Google's \tensorflow package and is capable of returning the probability for each pixel if it is contaminated with RFI.
The code is made publicly available under GPLv3 license and can be used for diverse generic classification problems.

We have made use of the open source \hide \& \seek radio data simulation and processing packages to train 
and assess the performance of our \unet. We first trained the network on a
simulated dataset where a perfect ground truth is available. We find that the \unet achieves an excellent result, confirming the applicability of CNN 
to this problem. We then applied the same code on radio data taken at the Bleien Observatory. As no ground 
truth is available with the observed dataset, we processed the TOD with \seek. \seek's 
\textsc{SumThreshold} RFI mitigation algorithm produced a RFI mask that we used as ground truth for training 
and performance measurement. The ground truth is however not perfect as it contains pixel that are incorrectly 
masked as RFI and some of the contaminated pixels were not flagged. We tested our \unet with various different 
configurations on the same data set, and all of them perform reasonably well in the receiver operating characteristic 
(ROC) and precision-recall curve. Both measures however, are strongly,
negatively biased as expected due to the imperfect ground truth. Visual
inspection of the predicted RFI mask shows that this approach is
showing competitive accuracy to classical RFI mitigation
algorithms such as \seek's \textsc{SumThreshold} implementation.

This work is the first step in applying CNN to the RFI mitigation problem. Several improvements and extensions 
are possible to make the method even more powerful. For example, one can
retrain the pre-trained network on a small data subset with an improved ground
truth. In addition, one can imagine altering the loss function so that masking the astronomical signal of interest is penalized. 
In light of the large amount of radio survey data that will become available in the coming decades, applications of 
the CNN techniques on RFI masking can potentially be a promising alternative to other conventional techniques. 

\section*{Acknowledgements}

We like to thank Lukas Gamper, Christian Monstein, Celine Blum and Adam Amara
for the useful discussions.

\appendix

\section{Distribution}
\label{sec:distribution}

The package is released under the GPLv3 license and the development is coordinated on GitHub 
\url{http://github.com/jakeret/tf_unet} and contributions are welcome.

To be able to use the package you have to clone the repository and make sure you have \tensorflow correctly 
installed on your machine\footnote{Installation instruction can be found here: \url{https://www.tensorflow.org/get_started/os_setup.html}}. 
All other project dependencies will be installed automatically during the setup procedure:

\begin{verbatim}
$ python setup.py develop --user
\end{verbatim}

\section{\tensorflow \unet usage example}
\label{sec:usage}

Our \tensorflow \unet implementation was written with flexibility in mind such that the package 
can be used for various classification problems. The typical use of our code looks like this:

\begin{verbatim}
from tf_unet import unet

net = unet.Unet(layers=3, features_root=64, channels=1, n_class=2)
trainer = unet.Trainer(net)
path = trainer.train(data_provider, output_path, training_iters=32, epochs=100)
\end{verbatim}

After importing the package we create a {\tt Unet} instance with 3 layers and 64 features. We use the network on 
grey-scaled images, hence we set {\tt channels} to 1. For RGB-images for example the number of channels would 
be 3. The network should learn to predict a binary problem (e.g. pixel is RFI or not). Therefore we set the number 
of classes to 2. Next, we create a {\tt Trainer} and train the network for 100 epochs with 32 training iterations. 
We pass an output path where the network and intermediate learning statistics should be stored, and a 
{\tt data\_provider} which can be a simple function or a callable that is providing data and labels for the training 
process. The code expects that the data has the shape {\tt [number of images, nx, ny, channels]} for the data and 
{\tt [number of images, nx, ny, number of classes]} for the one-hot encoded
labels, where {\tt nx} and {\tt ny} denotes the size of the images in pixels.

To obtain a prediction form the network we perform:

\begin{verbatim}
prediction = net.predict(output_path, x_test)
\end{verbatim}

We provide the path to the trained network on the file system and pass the data set on which we would like to run 
the network prediction. Both, the {\tt Unet} and the {\tt Trainer} implementation offer further parametrizations that 
are described in the package documentation.

\bibliographystyle{elsarticle-num}
\bibliography{rfi_unet}

\end{document}